\documentclass[aps,prl,twocolumn, superscriptaddress,floatfix]{revtex4}
\usepackage{bm}
\usepackage{graphicx}
\usepackage{ulem}
\def\up{\uparrow}
\def\down{\downarrow}
\def\be{\begin{equation}}
\def\ee{\end{equation}}
\def\ber{\begin{eqnarray}}
\def\eer{\end{eqnarray}}

\def\rv{{\bf r}}
\def\zv{{\bf z}}

\def\pv{{\bf p}}
\def\Vv{{\bf V}}

\def\jv{{\bf j}}
\def\kv{{\bf k}}
\def\qv{{\bf q}}

\def\Ev{{\bf E}}

\def\nn{\nonumber}

\begin{document}
\title{Spin Hall Drag}
\author{S. M. Badalyan}
\affiliation{Department of Physics, University of Regensburg, 93040 Regensburg, Germany and\\
Department of Radiophysics, Yerevan State University, 1 A. Manoukian St., Yerevan, 375025 Armenia }

\author{G. Vignale}
\affiliation{Department of Physics and Astronomy, University of Missouri, Columbia, Missouri 65211, USA}

\begin{abstract}
{We predict a new effect in electronic bilayers: the {\it Spin Hall Drag}.  The effect consists in the generation of spin accumulation across one layer by an electric current along the other layer.  It arises from the combined action of spin-orbit and Coulomb interactions.    Our theoretical analysis, based on the Boltzmann equation formalism,  identifies two main contributions to the spin Hall drag resistivity:  the side-jump contribution, which dominates at low temperature, going as $T^2$, and the skew-scattering contribution, which is proportional to $T^3$.  The induced spin accumulation is large enough to be detected in optical rotation experiments.}
\end{abstract} 
\maketitle

Double-layer structures consisting of two parallel quantum wells separated by a potential barrier are an important class of nanoscale electronic devices. Each layer hosts a quasi-two dimensional electron gas and electrons interact across the barrier via the Coulomb interaction. When an electric current is driven in one of the layers, the Coulomb interaction causes a charge accumulation in the other layer, in which the current flows. This phenomenon is known as {\it Coulomb drag} (CD)~\cite{Pogrebinskii77}-\cite{Rojo99} and is depicted in Fig. (1a).   The Coulomb drag resistivity $\rho_{CD}=E_{2x}/j_{1x}$ depends on the rate of momentum transfer between the layers and is largely independent of the scattering mechanism in each layer.   Because of  the requirements of momentum and energy conservation in electron-electron scattering $\rho_{CD}$ vanishes as $T^2$ at low temperature $T$.  A typical value in GaAs quantum wells is $\rho_{CD}\sim$20 ${\rm \Omega}$ at a temperature of a few Kelvin~\cite{bkvs2007,kellogg}.  

Another effect of great current interest is the {\it Spin Hall Effect}~\cite{Dyakonov71}-\cite{Stern06}, i.e. the generation of a transversal spin accumulation by an electric current in a single electron layer.  This effect, depicted in  Fig. (1b),  is due to spin-orbit interaction with impurities in a single electron layer.  The analysis of the effect is greatly simplified by considering quantum wells of special orientation relative to the crystallographic axes, e.g. [110] quantum wells in zincblende semiconductors such as GaAs.  In these quantum wells the component of the electron spin perpendicular to the plane  (hereafter denoted by $z$)  is essentially conserved, i.e., spin-flipping interactions are known to be weak.   Due to spin orbit coupling, electrons are preferentially scattered to the right or to the left of the impurity  according to their spin orientation.  This spin-biased scattering gives rise to  ``spin accumulation", i.e.  a gradient of spin electrochemical potential $E_{1\sigma y} = \sigma E_{1y}$ ($\sigma =+1$ or $-1$ for spin up and spin down respectively) in the direction perpendicular to the current.  The value of the spin Hall resistivity  $\rho_{SH,1}=E_{1y}/j_{1x}$  is weakly temperature dependent and is typically found to be a small fraction  ($10^{-3}$)  of the Drude resistivity\cite{Sih05,Hankiewicz06a,Tse06}.

In this article we predict and analyze theoretically a new effect arising from the combined action of spin-orbit interaction in the layers and Coulomb interaction between the layers. The effect consists in the generation of spin accumulation in one layer by an electric current in the other layer, and is depicted in Fig. (1c).  
\begin{figure}
\includegraphics[width=0.6\linewidth]{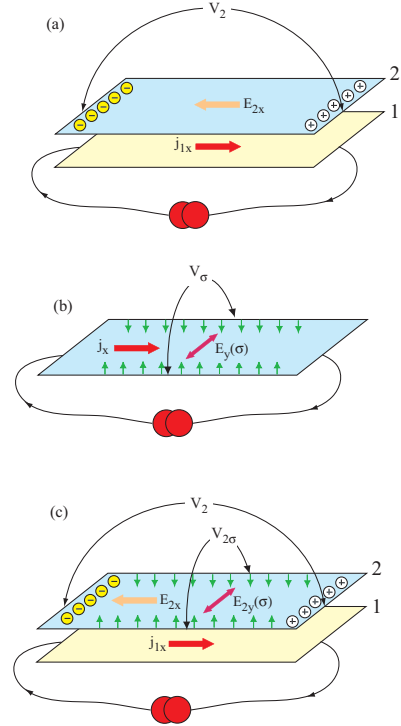}
\caption{(Color online) (a) In ordinary Coulomb drag the current $j_{1x}$ in layer 1 induces, via inter-layer Coulomb interaction, an electrochemical potential gradient $E_{2x}$ in layer 2.  (b) In spin Hall effect the current $j_{x}$ in a single layer induces, via spin-orbit interaction, a spin-dependent electrochemical potential gradient $E_{y}(\up) = -E_{y}(\down)$  causing electrons of opposite spin  orientation to accumulate on opposite edges.   (c) In Spin Hall drag the current $j_{1x}$ in layer 1 induces, via a combination of inter-layer Coulomb interaction and  spin-orbit interaction,  electrochemical potential gradients $E_{2x}$ {\it along} layer 2, and $E_{2y}(\sigma)$  {\it across} layer 2.}  
\label{CD-SHE}
\end{figure}
Because there is no current flowing in layer 2 there is no question of impurity scattering giving rise to an ordinary spin Hall effect in this layer.  However, we predict that a spin Hall accumulation, described by a gradient of spin electrochemical potential $E_{2\sigma y} = \sigma E_{2y}$ will still arise due to mechanisms that involve the Coulomb interaction between the two layers.   In the absence of intrinsic spin precession (the only case we consider here) there are two such mechanisms, skew-scattering and side-jump, and their relative importance will be discussed below.  Our calculations indicate that the induced spin accumulation is large enough to be detected in optical rotation experiments.\\

{\it Theory~} 
The linearized Boltzmann equation offers a convenient framework for analyzing the spin Hall drag.  For electrons in  layer 2  (the passive layer) we have
\be\label{BE}
-e\Ev_{2}(\sigma) \cdot \frac{\partial f^{(0)}_{2} (\epsilon_{\kv\sigma})}{\partial \kv}=I[f_{2\kv\sigma}]\,,
\ee
where  $f^{(0)}_{2} (\epsilon_{\kv\sigma})$ is the equilibrium distribution in layer 2, $\Ev_{2}(\sigma)$ is the gradient of electrochemical potential for spin-$\sigma$ and $I[f_{2\kv\sigma}]$ is the ``collision integral",  which includes both electron-impurity collisions in layer 2 and Coulomb collisions with electrons in layer 1.   The ``spin Hall drag accumulation"  is given by
\be
\Delta \mu_{SHD} = [E_{2y}(\up)-E_{2y}(\down)] w
\ee
where $w$ is the width of the layers.  The fields $\Ev_{2}(\sigma)$ are easily obtained from Eq.~(\ref{BE}) as
\be\label{BE2}
\Ev_{2}(\sigma) = \frac{1}{e n_{2\sigma} {\cal A}}\sum_\kv \kv I[f_{2\kv\sigma}]\,,
\ee
where $n_{i\sigma}$ is the electron density in layers $i$ and ${\cal A}$ is the area of each layer. 
\begin{widetext}
The collision integral is the sum of an electron-impurity term and an electron-electron term: $I=I^{ei}+I^{ee}$. The electron-impurity term is
\be\label{collision-ei}
I^{ei}[f_{2\kv\sigma}] = -\sum_{\kv'}\left(W^{ei}_{\kv \sigma,\kv'\sigma}f_{2\kv\sigma}-W^{ei}_{\kv' \sigma,\kv\sigma}f_{2\kv'\sigma}\right)\delta(\tilde\epsilon_{2\kv\sigma}-\tilde\epsilon_{2\kv'\sigma})\,,
\ee
where $W^{ei}_{\kv \sigma,\kv'\sigma}\delta(\tilde\epsilon_{2\kv\sigma}-\tilde\epsilon_{2\kv'\sigma})$ is the transition rate from $|2\kv\sigma\rangle$ to $|2\kv'\sigma\rangle$ under the influence of the electron-impurity potential.  Similarly, the electron-electron term is
\ber\label{collision-ee}
I^{ee}[f_{2\kv\sigma}]&=& -\sum_{\kv',\pv,\pv',\tau} \delta_{\kv+\pv,\kv'+\pv'} \left\{W^{ee}_{\kv\sigma,\pv\tau;\kv'\sigma,\pv'\tau}f_{2\kv\sigma}f_{1\pv\tau}(1-f_{2\kv'\sigma})(1-f_{1\pv'\tau})\right.\nn\\
&-&\left.  W^{ee}_{\kv'\sigma,\pv'\tau;\kv\sigma,\pv\tau}f_{2\kv'\sigma}f_{1\pv'\tau}(1-f_{2\kv\sigma})(1-f_{1\pv\tau})\right \}
\delta(\tilde\epsilon_{2\kv\sigma}+\tilde\epsilon_{1\pv\tau}-\tilde\epsilon_{2\kv'\sigma}-\tilde\epsilon_{1\pv'\tau})\,,\nn\\
\eer
where $W^{ee}_{\kv\sigma,\pv\tau;\kv'\sigma,\pv'\tau}\delta(\tilde\epsilon_{2\kv\sigma}+\tilde\epsilon_{1\pv\tau}-\tilde\epsilon_{2\kv'\sigma}-\tilde\epsilon_{1\pv'\tau})$ is the transition rate from  $|2\kv \sigma,1\pv \tau\rangle$ to $|2\kv'\sigma,1\pv'\tau\rangle$ under the influence of the interlayer Coulomb interaction. 
\end{widetext}
The spin-orbit interaction enters the above expressions in two distinct ways.  First, the conservation of energy is {\it not} formulated in terms of the ordinary energy $\epsilon_\kv = k^2/2m$, but in terms of the modified energy~\cite{Nozieres73,Hankiewicz06a}
\be\label{etilde}
\tilde \epsilon_{i\kv\sigma} \equiv \epsilon_\kv+2e\alpha\sigma  (\kv \times \Ev_{i}(\sigma))_z\,\,,
\ee
where $i=1,2$ denotes the layer and $\alpha$ is the spin-orbit coupling constant for the conduction band of the semiconductor ($\alpha\hbar  = 4.4 \times 10^{-20}$ m$^2$ in GaAs)~\cite{Winkler03}.   The reason for this is that the energy of an electron in the presence of the electric field is given by $\epsilon_\kv+e\Ev\cdot\rv+e\alpha\sigma  (\kv \times \Ev)_z$, and the last two terms in this expression change by equal amounts  during a collision process. The replacement of $\epsilon_\kv$ by $\tilde \epsilon_\kv$ is the mathematical expression of the ``side jump effect"~\cite{Culcer09}.    Second,  the scattering probabilities,  calculated beyond the first Born approximation but to first order in $\alpha$,  are not symmetric under interchange of the initial and final momenta.  Taking into account time-reversal invariance,  we can write
\be\label{WEI}
W^{ei}_{\kv\sigma,\kv'\sigma}=W^{ei,s}_{\kv,\kv'}+ \alpha \sigma W^{ei,a}_{\kv,\kv'}\,
\ee
where $W^{ei,s}_{\kv,\kv'}$ and $W^{ei,a}_{\kv,\kv'}$  are, respectively, symmetric and antisymmetric upon interchange of $\kv$ and $\kv'$: $W^{ei,s}_{\kv,\kv'} = W^{ei,s}_{\kv',\kv}$ and $W^{ei,a}_{\kv,\kv'} = -W^{ei,a}_{\kv',\kv}$. Similarly we can write
\ber\label{WEE}
W^{ee}_{\kv\sigma,\pv\tau;\kv'\sigma,\pv'\tau} &=& W^{ee,s1}_{\kv,\pv;\kv',\pv'}+\sigma \tau W^{ee,s2}_{\kv,\pv;\kv',\pv'} \nn\\&+&\frac{\alpha}{2} (\sigma+\tau)W^{ee,a}_{\kv,\pv;\kv',\pv'}\,,
\eer
where $W^{ee,s1}$ and $W^{ee,s2}$ are symmetric under interchange of the initial and final states and  $W^{ee,a}$ is antisymmetric: $W^{ee,a}_{\kv,\pv;\kv',\pv'} = -W^{ee,a}_{\kv',\pv';\kv,\pv}$.
The presence of the antisymmetric component $W^{ei,a}$ (Eq.~(\ref{WEI})) is responsible for the skew-scattering contribution to the ordinary Hall effect in layer 1.  And the presence of the antisymmetric component $W^{ee,a}$ (Eq.~(\ref{WEE}))  is responsible for the Coulomb skew scattering contribution to the spin Hall drag effect in layer 2.

Following the standard procedure for steady-state transport, we assume that the non-equilibrium distribution in layer 1  has the form of a shifted Fermi distribution
\be\label{f-Ansatz1}
f_{1\kv\sigma}=f^{(0)}_{1}(\epsilon_{\kv\sigma})-f^{(0)'}_{1} (\epsilon_{\kv\sigma}) \kv \cdot \Vv_{1}\,,
\ee
where $\Vv_{1}$ is the average drift velocity of electron gas in layer 1, and  $f^{(0)'}_{1}$ denotes the derivative of the equilibrium distribution with respect to energy.  At the same time we set
\be\label{f-Ansatz2}
f_{2\kv\sigma}=f^{(0)}_{2}(\epsilon_{\kv\sigma})\,,
\ee
meaning that the distribution of electrons in layer 2 remains unshifted from equilibrium, so that, in particular, the current is zero~\footnote{Although the expression of the current is modified by spin-orbit interaction, the corrections involve the net force on the electron and vanish in the steady state.}.

To first order in $\alpha$ the anomalous energy and the asymmetric scattering probability give independent contributions to the spin Hall drag accumulation, so we can study them separately.\\

{\it Coulomb side jump~}  To calculate the Coulomb side-jump contribution we treat the scattering probability to zero-th order in $\alpha$, so only its symmetric component survives. However, we retain the spin-orbit coupling terms in the  conservation of energy.  We rewrite the non-equilibrium distribution functions~(\ref{f-Ansatz1},\ref{f-Ansatz2})  as follows:
\ber\label{fnoneq}
f_{1\kv\sigma}&=&f^{(0)}_{1} (\tilde \epsilon_{1\kv\sigma})\nn\\
&-&f^{(0)'}_{1} (\epsilon_{\kv\sigma}) \left[\kv \cdot \Vv_{1}-2e\alpha\sigma  (\Ev_{1}(\sigma)\times \kv)_z \right]\,,\nn\\
f_{2\kv\sigma}&=&f^{(0)}_{2} (\tilde \epsilon_{2\kv\sigma})+f^{(0)'}_{2} \left[2e\alpha\sigma  (\Ev_{2}(\sigma)\times \kv)_z \right]\,.
\eer
The ``zero-th order terms",  $f^{(0)}_{i} (\tilde \epsilon_{\kv\sigma})$, are annihilated by the collision integral and can be discarded.  The remaining terms are of first order in the deviation from equilibrium and their contribution to the collision integrals~(\ref{collision-ei}-\ref{collision-ee}) can be calculated neglecting the difference between $\tilde \epsilon$ by $\epsilon$ in the $\delta$-function that expresses the conservation of energy.   A direct calculation  of the spin Hall drag resistivity gives
\be\label{MainResult}
\Ev^{sj}_{2}(\sigma) = -2 \sigma \rho_2  n_\sigma e^2\alpha  \rho_{CD}\jv_1\times \zv\,,
\ee
where $\rho_2$ is the Drude resistivity per spin channel in layer 2 and $\rho_{CD}$ is the Coulomb drag resistivity~\footnote{In arriving at Eq.~(\ref{MainResult}) we have neglected terms proportional to $(\rho_{CD}/\rho_2)^2$}.   The expression for $\rho_{CD}$ is well known (as is the fact that it vanishes at low temperature as $T^2$) and needs not be reproduced here.  
More important for the present discussion is the fact that the spin Hall drag resistivity  $\rho_{SHD} = |E_{2y}(\sigma)/j_{1x}|$ is related to the Coulomb drag resistivity by
\be
\rho_{SHD} = \frac{2 e \alpha}{\mu_2}\rho_{CD}\,,
\ee
where $\mu_2$ is the mobility of electrons in layer 2.   Notice that the resistivity is inversely proportional to $\mu_2$: thus the effect will be {\it larger} in low-mobility samples provided disorder is not so strong as to cause a breakdown of the Fermi liquid picture, e.g. localization.
In order to give a conservative estimate of  $\rho_{SHD}$ we assume  $\rho_{cd}$ =20~$\Omega$\footnote{The value of $\rho_{CD}$ is  found in GaAs quantum wells of width 18 nm, separated by a distance of 28 nm, at a sheet density of 3.8 $\times$ 10$^{14}$ m$^{-2}$ and a temperature of 5~K (see Ref.~\cite{bkvs2007}).} and $\mu_2$ =0.1~{\rm m$^2$/(V.s)}: then,  with $\alpha\hbar  = 4.4 \times 10^{-20}$ m$^2$, we obtain $\rho_{SHD} \simeq 0.026~\Omega$.   For a current density $j_{1x} \sim 1~ {\rm A/m}$ in the active layer this implies a spin-splitting of the chemical potential of about $5 \times 10^{-3}$ meV over a transverse width   $w= 100$~$\mu$m.   This splitting is about 200 times smaller than the splitting of approximately 1 meV previously observed in spin Hall effect measurements in GaAs quantum wells, but should be within the reach of modern spin detection techniques.\\  

{\it Coulomb skew scattering~}
To estimate the skew-scattering effect  we consider the contribution of the antisymmetric components of the scattering probabilities $W^{ei,a}$ and $W^{ee,a}$ to the collision integrals
~(\ref{collision-ei}-\ref{collision-ee}).  In this calculation the difference between $\tilde \epsilon$ and $\epsilon$ can be ignored.  It is readily seen that the electron-impurity skew scattering gives no contribution because there is no current in layer 2.   The  Coulomb skew scattering term can be expressed compactly under the assumption that $W^{ee,a}_{\kv,\pv;\kv',\pv'}$ depends only on the magnitude of the momentum transfer $q=|\qv|=|\kv'-\kv|=|\pv-\pv'|$, and on the sine of the angle between $\kv$ and $\kv'$, where both $|\kv|$ and $|\kv'|$ are close to the Fermi momentum $k_F$:
$ W^{ee,a}_{\kv,\pv;\kv',\pv'} = W^{ee,a}(q) (k_x q_y-k_yq_x)/k_F^2$.
A straightforward calculation leads to the formula
\ber\label{Skew-Coulomb2.4} 
E^{ss}_{2y}(\sigma)&=&-\frac{\hbar  j_{1x}}{e^2}\frac{\alpha \sigma}{64n_{2\sigma}}\int_0^\infty dq q  W^{ee,a} \nn\\
&&\int_{0}^{\infty}\frac{d\hbar\omega}{k_BT} \left(\frac{\hbar \omega}{2E_F}\right)^2 
\frac{S_0(q,\omega)\Gamma_0(q,\omega)}{\sinh^2 (\hbar\omega/2k_BT)}\,,\nn\\
\eer
 where the spectra $S_0(q,\omega)$ and $\Gamma_0(q,\omega)$ are defined as
 \be\label{S0}
S_0(q,\omega)= \sum_{\kv}(f^{(0)}_{2\kv}-f^{(0)}_{2{\kv+\qv}})  \delta(\epsilon_{\kv}-\epsilon_{\kv-\qv}-\omega)\ee
(the dynamical structure factor of the electron gas at zero temperature) and
\ber\label{Gamma0}
\Gamma_0(q,\omega) &=& \sum_{\pv,\tau}(f^{(0)}_{1\pv\tau}-f^{(0)}_{1{\pv-\qv}\tau}) \delta(\epsilon_{\pv}-\epsilon_{\pv-\qv}+\omega)  \nn\\
&\times&\left\{\left(\tanh \frac{\epsilon_p}{2k_BT}+\tanh \frac{\epsilon_{\pv-\qv}}{2k_BT}\right) \right.\nn\\
&+&\left. \frac{\hbar q^2}{2m\omega}\left(\tanh \frac{\epsilon_{\pv-\qv}}{2k_BT}-\tanh \frac{\epsilon_p}{2k_BT}\right)\right\}\,.\nn\\
\eer

 The important point is that $S_0(q,\omega)$ vanishes linearly with $\omega$ (independent of temperature), while $\Gamma_0(q,\omega)$ vanishes as $\hbar \omega/k_BT$ for $\omega \to 0$ ($\hbar\omega\ll k_BT$).  Since the $\sinh^2 (\hbar\omega/2k_BT)$ restricts the frequency integral in Eq.~(\ref{Skew-Coulomb2.4}) to $\hbar \omega \stackrel{<}{\sim} k_BT$ we can immediately conclude that the skew-scattering contribution to the resistivity vanishes as $T^3$ in the low-temperature limit.  A comparison between skew-scattering and side-jump contributions to the spin Hall drag resistivity is shown in the inset of Fig. 2.

\begin{figure}
\includegraphics[width=0.8\linewidth]{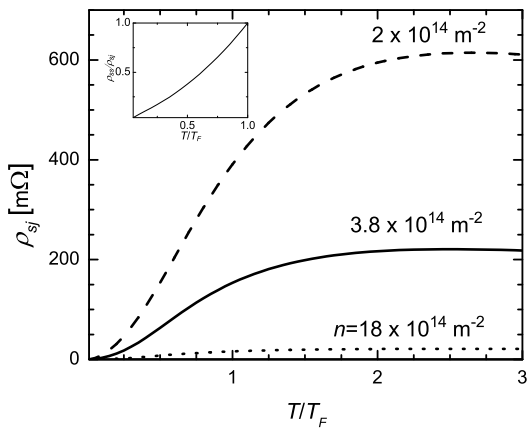}
\caption{Side-jump contribution to the spin Hall drag resistivity vs temperature $T/T_F$. The calculation includes dynamical screening, static exchange-correlation, and quantum well width effects along the lines of Ref.~\onlinecite{bkvs2007}.  The width of the quantum well is $18$ nm and the distance between the centers of the wells is $28$ nm.  The solid, dashed, and dotted lines correspond to the electron sheet densities  of $18 \times 10^{14}$ m$^{-2}$, $3.8 \times 10^{14}$ m$^{-2}$  and $2 \times 10^{14}$ m$^{-2}$.  The inset shows the ratio of the skew scattering resistivity, evaluated from Eq.~(\ref{Skew-Coulomb2.4}) with $W^{ee,a}$,  to the side-jump resistivity.  The value of $W^{ee,a}$ is chosen so that this ratio is $1$ at $T=T_F$.  The linear increase at low temperatures illustrates the $T^3$ behavior of skew-scattering resistivity, in contrast to the usual $T^2$ dependence of side-jump drag.} 
\label{Results}
\end{figure}

In summary, we have presented a theoretical analysis  of a new many-body effect in coupled bilayer systems: the spin Hall drag.  We have considered only the simplest situation, in which the so-called intrinsic spin Hall effect\cite{Sinova04} is absent.  Under these conditions we have identified the side-jump effect as the dominant contribution to the spin Hall drag resistivity, varying as $T^2$  in the low-temperature Fermi liquid regime.  By contrast, the Coulomb skew-scattering mechanism vanishes  as $T^3$.  From a theoretical point of view it is remarkable that the two contributions are  distinguished by different temperature dependences.  From an experimental point of view, the spin Hall drag accumulation can be measured by optical rotations techniques, which do not require the fabrication of separate electrical contacts for layer 2.  Our numerical estimates indicate that the prospects for observation of the extrinsic effect are quite good.   By experimenting on samples grown in different directions it should also be possible to study the interplay between intrinsic and extrinsic contributions to the spin Hall drag.
 
We thank N. Samarth, D. Awschalom, and J. Fabian for useful discussions.  G.V. acknowledges support from NSF Grant No. 0705460  and from the Ikerbasque Foundation at the ETSF in San Sebasti\'an.  S.M.B. acknowledges support from EU Grant PIIF-GA-2009-235394, SFB Grant 689, and ANSEF Grant PS-1576.

\end{document}